\documentclass[twocolumn]{revtex4}
\usepackage[english]{babel}
\usepackage{amsfonts}
\usepackage{amssymb}
\usepackage[dvips]{graphicx}
\usepackage{color}

\begin{document}

\title{ Experimental verification of a modified fluctuation-dissipation relation
for a micron-sized particle in a non-equilibrium steady state}
\author{J. R. Gomez-Solano, A. Petrosyan, S. Ciliberto, R. Chetrite, and K.
Gaw\c{e}dzki} \affiliation{ Universit\'e de Lyon, Laboratoire de
Physique, Ecole Normale Sup\'erieure de  Lyon, CNRS, 46, All\'ee
d'Italie, 69364 Lyon CEDEX 07, France}

\begin{abstract}
A modified fluctuation-dissipation-theorem (MFDT) for
a non-equilibrium steady state (NESS) is experimentally checked by
studying the position fluctuations of a colloidal particle whose
motion is confined in a toroidal optical trap. The NESS is
generated by means of a rotating laser beam which exerts on the
particle a sinusoidal conservative force plus a constant
non-conservative one. The MFDT is shown to be perfectly verified
by the experimental data. It can be interpreted as an
equilibrium-like fluctuation-dissipation relation in the Lagrangian
frame of the mean local velocity of the particle.
\end{abstract}


\maketitle

The validity of the fluctuation-dissipation theorem (FDT) in systems out of thermal equilibrium
has been the subject of intensive study during the last years. We recall that for a system
in equilibrium with a thermal bath at temperature $T$
the FDT establishes a simple relation between the 2-time correlation function
$C(t-s)$ of a given observable and the linear response function $R(t-s)$ of this observable to a
weak external perturbation
\begin{equation}\label{eq:FDT}
        \partial_s C(t-s)=k_B T R(t-s).
\end{equation}
However, Eq.~(\ref{eq:FDT}) is not necessarily fulfilled out of equilibrium and violations
are observed in a variety of systems such as glassy materials \cite{cugliandolo,grigera,berthier,crisanti,calabrese},
granular matter \cite{barrat}, and biophysical systems \cite{hayashi1}.
This motivated a theoretical work devoted to a search of a general framework
describing FD relations, see the review \cite{marconi}
or \cite{hayashi2,harada,speck1,bickle1,sakaue,baiesi}
for recent attempts in simple stochastic systems.
In the same spirit, a modified fluctuation-dissipation theorem (MFDT) has been
recently formulated for a non-equilibrium steady dynamics
governed by the Langevin equation with non-conservative forces \cite{chetrite}.
In particular, this MFDT holds for the overdamped motion of a particle
on a circle, with angular position $\theta$, in the presence of a
periodic potential
$H(\theta)=H(\theta+2\pi)$ and a constant non-conservative force $F$
\begin{equation}\label{eq:Langevin}
        \dot{\theta} = - \partial_{\theta}H(\theta) + F + \zeta,
\end{equation}
where $\zeta$ is a white noise term of mean $\langle \zeta_t
\rangle= 0$ and covariance $\langle \zeta_t\zeta_s \rangle = 2 D
\delta(t-s)$, with $D$ the (bare) diffusivity.
This is a system that may exhibit an increase
in the effective diffusivity \cite{reimann,lee}. Here, we shall study
the dynamical non-equilibrium steady state (NESS) reached for observables
that depend only on the particle position on the circle so are periodic
functions of the angle $\theta$. Such a state corresponds to a constant
non-vanishing probability current $j$ along the circle and a periodic
invariant probability density function $\rho_0(\theta)$ that allow
us to define a mean local velocity $v_0(\theta) = j/\rho_0(\theta)$. This
is the average velocity of
the particle at $\theta$. For a stochastic system in
NESS evolving according to Eq.~(\ref{eq:Langevin}), the MFDT reads
for $t \ge s$
\begin{equation}\label{eq:MFDT}
        \partial_s C(t-s) - b(t - s) = k_B T R(t-s),
\end{equation}
where the 2-time correlation of a given observable $O(\theta)$ is defined by
\begin{equation}\label{eq:autocorrelation}
        C(t-s) = \langle \, O(\theta_t)O(\theta_s) \, \rangle_0,
\end{equation}
and the linear response function to a $\delta$-perturbation of
the conjugated variable $h_t$ is given by the functional derivative
\begin{equation}\label{eq:response}
        R(t-s)= \left. \frac{\delta}{\delta h_s}\right|_{h=0} \langle \, O(\theta_t) \, \rangle_h.
\end{equation}
In Eq.~(\ref{eq:response}), $\langle ... \rangle_h$ denotes the average
in the perturbed time-dependent state obtained from the NESS by replacing
$H(\theta)$ in Eq.~(\ref{eq:Langevin}) by $H(\theta)-h_tO(\theta)$. It
reduces for $h=0$ to the NESS average $\langle ... \rangle_0$.
In Eq.~(\ref{eq:MFDT}), the correlation $b(t-s)$ is given by
\begin{equation}\label{eq:corrective}
        b(t-s)= \langle \, O(\theta_t) {v_0}(\theta_s) \partial_{\theta}O(\theta_s) \, \rangle_0.
\end{equation}
This new term takes into account
the extent of the violation of the usual fluctuation-dissipation relation
(\ref{eq:FDT}) due to the probability current and it plays the role of a
corrective term to $C(t-s)$ in the  MFDT, Eq.~(\ref{eq:MFDT}),
which can be rewritten in the integral form:
\begin{equation}\label{eq:integMFDT}
       C(0)-C(t)-B(t)=k_B T \chi(t),
\end{equation}
where $B(t)\equiv \int_0^t b(t-s) ds$ and  $\chi(t)=\int_0^t
R(t-s) ds$ is the integrated response function.

In this letter, we  present an experimental test of
Eq.~(\ref{eq:integMFDT}) in the linear response regime around a
NESS attained by a micron-sized
particle in a toroidal optical trap similar to the one used in
\cite{bickle1}. We first show that the dynamics of the particle is
well described by the Langevin equation (\ref{eq:Langevin}) on a circle.
Secondly, by measuring $v_0$, $B(t)$, $C(t)$ and $\chi(t)$,
we verify Eq.~(\ref{eq:integMFDT}) for the observable $O(\theta)=\sin \theta$.
The result can be interpreted as
an equilibrium-like fluctuation-dissipation relation
in the Lagrangian frame of the mean
local velocity $v_0(\theta)$ \cite{chetrite}.
We also check that $\rho_0(\theta)$ is frame invariant.

\begin{figure}%[htp]
 \centering        {\includegraphics[width=.45\textwidth]{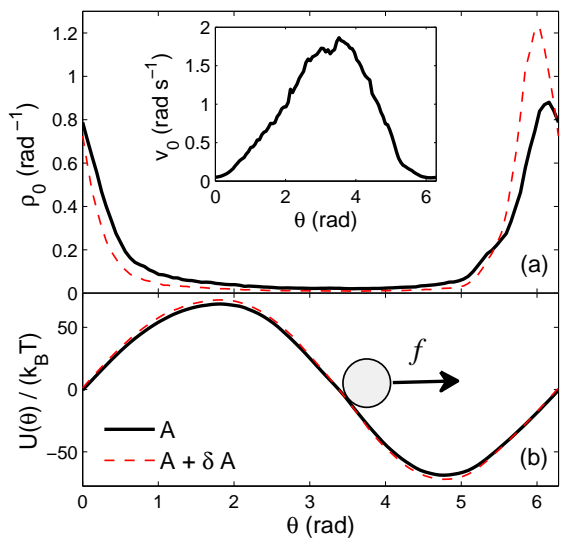}}
     \caption{(a) Invariant density of
     the angular position of the particle in NESS for a modulation of 7\%
     (black solid line) and 7.35\% (red dashed line) around the laser mean power.
     Inset: Mean local velocity of the particle in the former case.
     (b) Corresponding potential profiles. The arrow indicates the direction of the
     non-conservative force $f$.}
  \label{multifig1}
\end{figure}

The experiment is performed  using a spherical silica particle of
radius $r=$1 $\mu$m in ultrapure water at room temperature $T =
20.0 \pm 0.5^{\circ}$C at which the dynamic viscosity of water is
$\eta = (1.002 \mp 0.010) \times 10^{-3}$ Pa s. The
particle is kept  by an optical tweezers in a toroidal optical
trap. This kind of trap consists on a Nd:YAG diode pumped solid
state laser beam ($\lambda=$ 1064 nm) which is focused by a
microscope objective ($63\times$, NA = 1.4) and  scans (by means of
two acousto-optic deflectors) a circle of radius $a = 4.12$ $\mu$m
in the horizontal plane at a rotation frequency of 200 Hz.
The toroidal trap is created 10 $\mu$m above the inner bottom
surface of the cell where hydrodynamic boundary-coupling effects
on the particle are negligible. At a rotation frequency of 200 Hz,
the laser beam is not able to hold the particle but drags it
regularly a small distance on the circle when passing through it
\cite{faucheux}. The diffusive motion of the particle along the
radial and vertical directions during the absence of the beam
is less than 40 nm, thus  the angular position of
the particle $\theta$ (measured modulo $2\pi$) is the only
relevant degree of freedom of the dynamics. The laser power is
sinusoidally modulated around 30 mW with an amplitude of 7\% of
the mean power, synchronously with the deflection of the beam at
200 Hz creating a static sinusoidal intensity profile along the
circle. This trapping situation acts as a constant
non-conservative force $f$ associated to the mean kick which
drives the particle across a sinusoidal potential $U(\theta)$ due
to the periodic intensity profile.
%Images of the intensity contrast of the
%particle on the horizontal focal plane $x-y$ are recorded
%with a resolution of $160 \times 130$
%pixels at a sampling rate of 150 frames per second.
The particle barycenter $(x_t,y_t)$ is measured by
image analysis with an accuracy of the order of 1 nm at a sampling rate
of 150 Hz and exposure time of 1/300 s.
This measure allows us  to determine the angular
position of the particle $\theta_t$ with respect to the trap center.
For more details about the experimental apparatus
see Ref.~\cite{jop}.
We determine the value of $f$ and the profile of
$U(\theta)$ by means of the method described in \cite{bickle2}.
This method exploits the probability current $j$ and the invariant
density $\rho_0(\theta)$ in NESS to reconstruct the actual energy
landscape of the particle. We recorded 200 time series
$\{\theta_t\}$ of duration 66.67 s with different initial
conditions \{$\theta_0$\} sampled every 5 minutes in order to
measure $j$ and $\rho_0(\theta)$. The probability current is
related to the global mean velocity of the particle by the
expression $j = \langle \dot{\theta} \rangle_0 / (2\pi)$. The
value of $\langle \dot{\theta} \rangle_0$ is calculated from the
slope of the linear fit of the mean angular position of the
particle (not taken modulo $2\pi$) as a function of time leading
to $j = 3.76 \times 10^{-2}$ s$^{-1}$ in the direction of the
laser beam rotation. The invariant density, shown as a solid black line in
Fig.~\ref{multifig1}(a), is computed from the histogram of each time
series $\{\theta_t\}$ averaged over the 200 different initial
conditions. In Fig.~\ref{multifig1}(a) we also show the mean local
velocity $v_0(\theta)=j/\rho_0(\theta)$ of the particle.
From these quantities we obtain
$f = 3\eta r a j \int_0^{2\pi} \rho_0(\theta')^{-1} d\theta'
=6.60 \times 10^{-14}$ N
and $U(\theta)=-k_B T \log \rho_0(\theta)+\int_0^{\theta}(f-6\pi \eta r a j\rho_0(\theta')^{-1})a d\theta'
= A \sin \theta$ with amplitude $A = 68.8 k_B T$.
The experimental potential profile is
shown in Fig.~\ref{multifig1}(b) (black solid line). Hence,
the time evolution of $\theta$ is claimed to follow the Langevin
dynamics of Eq.~(\ref{eq:Langevin}) \cite{bickle2} with $F =
f/(6\pi \eta r a) = 0.85$ rad s$^{-1}$, $H(\theta)=U(\theta)/(6
\pi \eta r a^2)= H_0 \sin \theta$, $H_0 = A/(6 \pi \eta r a^2) =
0.87$ rad s$^{-1}$, and $D = k_B T/(6\pi \eta r a^2)=1.26\times
10^{-2}$ rad$^2$ s$^{-1}$.
Note that in the corresponding equilibrium situation
($f=0$) the probability maximum would be located
at the minimum of $U(\theta)$ ($\theta = 3\pi /2$). However, in NESS
the non-conservative force $f > 0$ shifts the maximum of
$\rho_0(\theta)$ in the positive direction,
as shown in Fig.~\ref{multifig1}(a).
The choice of the parameters has been done to enhance
the stochastic nature of the dynamics, i.e. we take
$(F-H_0)/H_0\approx-2\%$ which is close to the maximum increase 
of the effective diffusivity following Refs. \cite{reimann,lee}.

Additionally, 500 times series of duration 100 s were specially
devoted for the determination of $\chi(t)$. In this case, during
each interval of 100 s we apply from time $t_0$ to $t_0+\Delta t$
with $0<t_0<66.67$ s and $\Delta t=33.33$ s a step perturbation
changing the value of $A$ to $A+\delta A$. This is accomplished by
suddenly switching the laser power modulation from 7\% to 7.35\%
of the mean power (30 mW).
The experimental value of the amplitude perturbation
($\delta A = 0.05A$) is determined from independent NESS measurements
of $\rho_0(\theta)$ and $U(\theta)$ for a power modulation of 7.35\%
(shown in Figs.~\ref{multifig1}(a) and \ref{multifig1}(b) respectively
as red dashed lines) as described previously.
By keeping constant the mean power
during the switch we ensure that the value of $f$ remains also
constant, compare to a different time-dependent protocol explored
recently in \cite{bickle3}.
In this way, we extract 500
\emph{perturbed} trajectories $\{\theta_t\}_{\delta A}$ of
duration $\Delta t=33.33$ s. We ensure that after switching off
the perturbation the system actually has attained a NESS before
the beginning of the next step perturbation.

\begin{figure}%[htp]
\centering {\includegraphics[width=.45\textwidth]{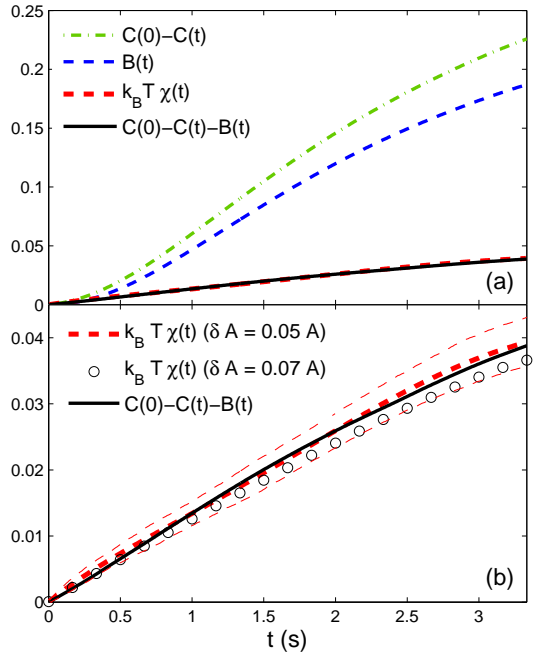}}
     \caption{(a) Comparison between the different terms needed to verify
     Eq.~(\ref{eq:integMFDT}), as functions of the time lag $t$.
     (b) Expanded view of the comparison between the curves $C(0)-C(t)-B(t)$
     and $k_B T \chi(t)$ shown in Fig \ref{multifig2}(a).
     The thin red dashed lines represent
     the error bars of the measurement of the integrated response.}
  \label{multifig2}
\end{figure}

\begin{figure}%[htp]
 \centering        {\includegraphics[width=.45\textwidth]{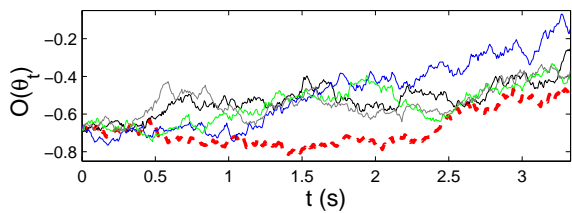}}
     \caption{Example of trajectories used to compute the integrated response.
     We show a perturbed trajectory (thick dashed red line) and four out of a total of
     200 of the corresponding unperturbed ones (see text).}
  \label{fig:trajectories_response}
\end{figure}

With the purpose of determining correctly the different terms involved 
in Eq.~(\ref{eq:integMFDT}), the observable $O(\theta)$ must be chosen 
consistently on both sides of such relation. The change of the potential 
$U(\theta) \rightarrow U(\theta)+\delta A \sin\theta$ due to the application
of the step perturbation implies that $O(\theta) = \sin \theta$ is the 
observable that must be studied with $-\delta A$ as its conjugate variable. 
Hence, we compute the correlation function $C(t)$, the corrective term $B(t)$ 
(with $\partial_{\theta} O(\theta) = \cos \theta$ and the experimental curve 
$v_0(\theta)$ shown in Fig.~\ref{multifig1}(a)) and the integrated response 
$\chi(t)$ for this observable, as functions of the time lag $t$.

The determination of $C(t)$ and $B(t)$ is straightforward
according to Eqs.~(\ref{eq:autocorrelation}) and
(\ref{eq:corrective}). The stationarity of the system allows
us to perform an average over the time origin in
addition to the ensemble average $\langle ... \rangle_0$ over the
200 different time series devoted to this purpose, which increases
enormously the statistics. The dependence of $C(0)-C(t)$ and
$B(t)$ on $t$ is shown in Fig. \ref{multifig2}(a) in green
dotted-dashed and blue dashed lines, respectively.

On the other hand, the integrated response $\chi(t)$ is given by
\begin{equation}\label{eq:integresponse}
    \chi(t)=\frac{\langle \, O(\theta_t) \, \rangle_{\delta A} - \langle \, O(\theta_t) \, \rangle_{0}}{-\delta A}.
\end{equation}
In Eq.~(\ref{eq:integresponse}) the value $t=0$ corresponds to
instant $t_0$ when the perturbation of the potential amplitude
$\delta A$ is switched on.
To decrease the statistical errors in comparison of the terms
in Eq.~(\ref{eq:integresponse}), for a given perturbed
trajectory ${\theta_t}_{\delta A}$ we look for as many unperturbed ones
$\theta_t$ as possible among the 200 time series $\{ \theta_t \}$ starting at
a time $t^*$ such that $O(\theta_{t^*})=O({\theta_0}_{\delta A})$.
Then we redefine $t^*$ as $t=0$ in Eq.~(\ref{eq:integresponse}),
as shown in Fig.~\ref{fig:trajectories_response}.
The unperturbed trajectories
found in this way allow us to define a subensemble over which the
average $\langle O(\theta_t) \rangle_0$ in
Eq.~(\ref{eq:integresponse}) is computed at a given $t$. The
average $\langle O(\theta_t) \rangle_{\delta A}$ is simply
computed over the 500 perturbed time series. In Fig.
\ref{multifig2}(a) we show in thick dashed red line the dependence
of the integrated response on $t$.

The comparison between the different terms needed to verify
Eq.~(\ref{eq:MFDT}) is shown in Fig.~\ref{multifig2}(a), for the
time lag interval $0<t<3.5$ s.
As expected, the usual FD relation (\ref{eq:FDT}) is strongly
violated in this NESS because of the broken detailed balance, with
the correlation term $C(0)-C(t)$ being one order of magnitude
larger than the response term $k_B T \chi(t)$. However, with the
corrective term $B(t)$ associated to the probability current
subtracted, $C(0)-C(t)-B(t)$ shown in solid black line in
Fig.~\ref{multifig2}(a), becomes equal to $k_B T \chi(t)$. For
clarity, in Fig. \ref{multifig2}(b) we show an expanded view of
the of the curves $C(0)-C(t)-B(t)$ and $k_B T \chi(t)$. We observe
that, within the experimental error bars, the agreement between
both terms is quite good, verifying the integrated form of the
modified FD relation (\ref{eq:integMFDT}). The error bars of the
integrated response curve at each time lag $t$ are obtained from
the standard deviation of the subensemble of unperturbed
trajectories found for each perturbed trajectory, like the ones
shown in thin solid lines in Fig. \ref{fig:trajectories_response}.
We checked that the perturbation $\delta A = 0.05
A$ is small enough to remain within the linear response regime.
%for time lags $0 \le t \le$ 3.5 s.
This is quantitatively seen in Fig.~2(b) where the response
measured at $\delta A= 0.07A$ is represented by circles. We see
that $\chi(t)$ is independent of $\delta A$ within  experimental
errors  showing that we are in the linear response regime.
The MFDT is checked  only for the first 3.5 s because after this 
time the evaluation of $\chi(t)$ is affected by large errors. Indeed
during the measurement of $\chi(t)$, for finite $\delta A$ the
system is approaching a new NESS which depends non-linearly on
$\delta A$ (see the strong non-linear dependence of $\rho_0$ on
$\delta A$ in Fig.~\ref{multifig1}(a)). The perturbed trajectories 
diverge with respect to the
unperturbed ones (Fig. 3) in such a way that after 3.5 s
the non-linear effects and the statistical error bars of
$\chi(t)$ due to finite sampling become large.

\begin{figure}%[htp]
 \centering        {\includegraphics[width=.435\textwidth]{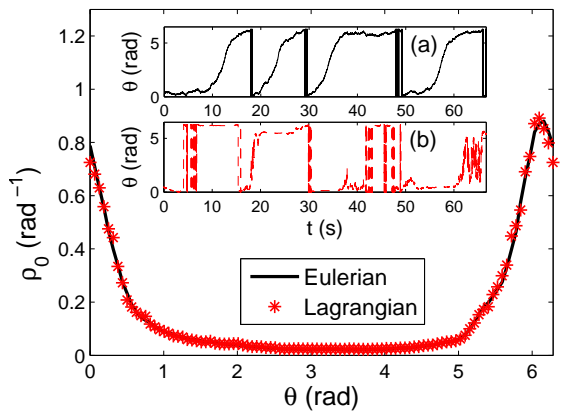}}
     \caption{Invariant density of the angular position measured in the
     Eulerian frame (continuous line)  (same as in Fig.~\ref{multifig1}(a)) and in
     the Lagrangian frame ($*$). Inset : example of a trajectory
     measured
     %measured
     respectively in the Eulerian (a) and
     Lagrangian (b) frames.}
  \label{fig:pdf_lagrangian}
\end{figure}

As shown in \cite{chetrite}, the validity of
Eq.~(\ref{eq:integMFDT}) for the fluctuations of the angular
position of the silica particle in NESS gains a simpler
interpretation in the Lagrangian frame of the mean local velocity
$v_0(\theta)$ along the circle.

Indeed, using the observables that are time independent in the
Lagrangian frame, the MFDT may be rewritten in the form
\begin{equation}
\partial_s C_L(t,s)=k_BT R_L(t,s),
\label{FDT_Lagrangian}
\end{equation}
 where $C_L$ and $R_L$ are the correlation and the response
measured in the Lagrangian frame. Eq.~(\ref{FDT_Lagrangian}) is
close to that of the equilibrium FDT (Eq.~(\ref{eq:FDT})) except
for the lack of the time translation invariance of the functions
involved. One of the new predictions of the Lagrangian analysis of
the system is that, although the trajectories in the Eulerian and
the Lagrangian frame are quite different, their average
density $\rho_0$ is the
same in the two frames. This property  is clearly illustrated by
the experimental data in Fig.~\ref{fig:pdf_lagrangian}, where we
compare the  densities measured in the two frames. The insets of
Fig. \ref{fig:pdf_lagrangian} point out to the difference
between a trajectory measured in the Eulerian frame and
the same trajectory measured in the Lagrangian frame.

We have verified experimentally a modified fluctuation-dissipation
relation describing the dynamics of a system with one degree of
freedom in NESS, namely a Brownian particle moving in a toroidal
optical trap. We point out that the experimental results reported
here represent an alternative approach to non-equilibrium
fluctuation-dissipation relations to that of Ref.~\cite{bickle1}
which dealt with the velocity fluctuations relative to the mean local
velocity. The approach followed in our work relies on an observable depending
on the particle position. It quantifies the extent of the violation
of the usual FDT by means of the term $B(t)$, admitting a transparent
Lagrangian interpretation of the resulting MFDT.


\begin{thebibliography}{99}

        \bibitem{cugliandolo} L. F. Cugliandolo, J. Kurchan, and L. Peliti, Phys. Rev. E \textbf{55}, 3898 (1997).
            \bibitem{grigera} T. S. Grigera and N. E. Israeloff, Phys. Rev. Lett., {\bf
83}, 5038 (1999); L. Bellon, S. Ciliberto, and C. Laroche, Europhys. Lett., {\bf 53} (4), 511 (2001);
D. Herisson and M. Ocio, Phys. Rev. Lett. \textbf{88}, 257202 (2002); L. Buisson and S.
Ciliberto, Physica D, {\bf 204} (1-2) 1 (2005).
            \bibitem{berthier} L. Berthier and J.-L. Barrat, Phys. Rev. Lett. \textbf{89}, 095702 (2002).
            \bibitem{crisanti} A. Crisanti and F. Ritort, J. Phys. A \textbf{36}, R181 (2003).
            \bibitem{calabrese} P. Calabrese and A. Gambassi, J. Phys. A \textbf{38}, R133 (2005).
            \bibitem{barrat} A. Barrat, V. Colizza, and V. Loreto, Phys. Rev. E \textbf{66}, 011310 (2002).
            \bibitem{hayashi1} K. Hayashi and M. Takano, Byophys. J. \textbf{93}, 895 (2007).
            \bibitem{marconi} U. Marini Bettolo Marconi, A. Puglisi, L. Rondoni and A. Vulpiani,
Physics Reports \textbf{461}, 111 (2008).
            \bibitem{hayashi2} K. Hayashi and S. I. Sasa, Phys. Rev. E \textbf{69}, 066119 (2004).
            \bibitem{harada} T. Harada and S. I. Sasa, Phys. Rev. Lett. \textbf{95}, 130602 (2005).
            \bibitem{speck1} T. Speck and U. Seifert, Europhys. Lett. \textbf{74}, 391 (2006).
            \bibitem{bickle1} V. Blickle, T. Speck, C. Lutz, U. Seifert, and C. Bechinger,
  Phys. Rev. Lett. \textbf{98}, 210601 (2007).
            \bibitem{sakaue} T. Sakaue and T. Ohta, Phys. Rev. E {\bf 77}, 050102(R) (2008).
        \bibitem{baiesi} M. Baiesi, C. Maes, and B. Wynants, arXiv:0902.3955v1 [cond-mat.stat-mech]
            \bibitem{chetrite} R. Chetrite, G. Falkovich, and K. Gawedzki, J. Stat. Mech.
    P08005 (2008).
            \bibitem{reimann} P. Reimann, C. Van den Broeck, H. Linke, P. H\"{a}nggi, J. M. Rubi, and A. P\'erez-Madrid, Phys. Rev. Lett. \textbf{87}, 010602 (2001).
            \bibitem{lee} S.-H. Lee, and D. G. Grier, Phys. Rev. Lett. \textbf{96}, 190601 (2006).
            \bibitem{faucheux} L. P. Faucheux, G. Stolovitzky, and A. Libchaber, Phys. Rev. E \textbf{51}, 5239 (1995).
            \bibitem{jop} P. Jop, J. R. Gomez-Solano, A. Petrosyan, and S. Ciliberto, J. Stat. Mech. P04012 (2009).
        \bibitem{bickle2} V. Blickle, T. Speck, U. Seifert, and C. Bechinger,
  Phys. Rev. E \textbf{75}, 060101(R) (2007).
        \bibitem{bickle3} V. Blickle, J. Mehl, and C. Bechinger, arXiv:0902.2650v1 [cond-mat.soft].

\end{thebibliography}
\end{document}